\documentclass{llncs}

\usepackage[dvips]{graphicx}
\usepackage{amsfonts}
\usepackage{amsmath}
\usepackage{latexsym}
\usepackage{amssymb}
\usepackage[mathcal]{euscript}

\newcommand{\ket}[1]{| #1 \rangle}
\newcommand{\bra}[1]{\langle #1 |}

\begin{document}


\title{The physical Church-Turing thesis and the principles of quantum theory}
\author{Pablo Arrighi\inst{1,2} \and Gilles Dowek\inst{3}}
\institute{
\'Ecole normale sup\'erieure de Lyon, LIP, 46 all\'ee d'Italie, 69008 Lyon, France\\
\and
Universit\'e de Grenoble, LIG, 220 rue de la chimie, 38400 SMH, France\\
\email{parrighi@imag.fr}
\and
INRIA\\
\email{Gilles.Dowek@inria.fr}
}
\maketitle

\begin{abstract}
Notoriously, quantum computation shatters complexity theory,
but is innocuous to computability theory \cite{Deutsch}.  Yet several
works have shown how quantum theory as it stands could breach the
physical Church-Turing thesis \cite{NielsenComputability,Kieu} We draw
a clear line as to when this is the case, in a way that is inspired by
Gandy \cite{Gandy}. Gandy formulates postulates about physics, such as
homogeneity of space and time, bounded density and velocity of
information --- and proves that the physical Church-Turing thesis is a
consequence of these postulates. We provide a quantum version of the
theorem. Thus this approach exhibits a formal non-trivial interplay
between theoretical physics symmetries and computability assumptions.
\end{abstract}

\section{Introduction}

The physical Church-Turing thesis states that any function that can be
computed by a physical system can be computed by a Turing
Machine. There are many mathematical functions that cannot
be computed on a Turing Machine (the halting function $h:{\mathbb
N}\to \{ 0,1 \}$ that decides whether the $i^{th}$ Turing Machine halts,
the function that decides whether a multivariate polynomial has
integer solutions, etc.). Therefore, the physical Church-Turing thesis
is a strong statement of belief about the limits of both physics and
computation.

The shift from classical to quantum computers challenges 
the notion of complexity: some functions can be computed faster 
on a 
quantum computer than on a classical one. But,
as noticed by Deutsch \cite{Deutsch}, 
it does not challenge 
the physical Church-Turing thesis itself:
a quantum computer can always be (very inefficiently) simulated
by pen and paper, through matrix multiplications. Therefore, what they 
compute can be computed classically.

Yet several researchers \cite{NielsenComputability,Kieu,NielsenMore}
have pointed out that Quantum theory does not forbid, in principle,
that some evolutions would break the physical Church-Turing
thesis. Indeed, if one follows the postulates by the book, the only
limitation upon evolutions is that they be unitary operators. Then,
according to Nielsen's argument \cite{NielsenComputability}, it
suffices to consider the unitary operator $U=\sum\ket{i,h(i)\oplus
b}\bra{i,b}$, with $i$ over integers and $b$ over $\{0,1\}$, to have a
counter-example. 

The paradox between Deutsch's argument and Nielsen's argument is only
an apparent one; both arguments are valid; the former applies
specifically to Quantum Turing Machines, the latter applies to
full-blown quantum theory. Nevertheless, this leaves us in a
unsatisfactory situation: if the point about the Quantum Turing
Machine was to capture Quantum theory's computational power, then it
falls short of this aim, and needs to be amended! Unless Quantum
theory itself needs to be amended, and its computational power brought
down to that of the Quantum Turing Machine?

Quantum theory evolutions are about possibly infinite-dimensional
unitary operators and not just matrices --- for a good reason: even
the state space of a particle on a line is infinite-dimensional. Can this
fact be reconciled with the physical Church-Turing thesis, at least at
the theoretical-level? Mathematically speaking, can we allow for all
those unitary operators we need for good physical reasons {\em and at
the same time} forbid the above $U=\sum\ket{i,h(i)\oplus b}\bra{i,b}$,
but for good physical reasons as well? These are the sort of questions
raised by Nielsen \cite{NielsenComputability}, who calls for a
programme of finding the non-ad-hoc, natural limitations that one
could place upon Quantum theory in order to make it computable: we
embark upon this programme of a computable Quantum theory.

The idea that physically motivated limitations lead to the physical
Church-Turing thesis has, in fact, already been investigated by Gandy
\cite{Gandy}.  Although some similarities exist, Gandy's proof of the
Church-Turing thesis serves different goals from those of the proof by Dershowitz and Gurevich
\cite{DershowitzGurevich}, as it is based not on an axiomatic notion of algorithm, but on physical hypotheses. 
In Gandy's proof, one finds the important idea that causality (i.e. bounded velocity of
information), together with finite density of information, could be
the root cause of computability (i.e. the physical Church-Turing
thesis). More generally, Gandy provides a methodology whereby
hypotheses about the real world have an impact upon the physical
Church-Turing thesis; an idea which can be transposed to other
settings: we transpose it to Quantum theory.

\section{Gandy's theorem for classical physics}
\label{sec:classic}

We first recall Gandy's argument in the classical case
\cite{Gandy,CopelandShagrir,SiegByrnes}.

We consider the tridimensional euclidean space $E$.  A {\em region} is
any subset of $E$. If $A$ is a region, we write $\Sigma(A)$ for the
set of possible states of $A$.  If $A$ is a region and $t$ a point in
time, we write $\rho(A,t)$ for the state of $A$ at time $t$. The state
$\rho(A,t)$ is an element of $\Sigma(A)$. For instance $\rho(E,t)$ is
the global state at time $t$, an element of $\Sigma(E)$ the global
state space.  A region is said to be {\em of finite size}, if it is
included in a sphere.  Let $A$ be a region, the {\em area of radius
$r$ around $A$} is the union of the closed spheres of radius $r$
centered on a point of $A$.

Gandy's hypotheses are the following.

\begin{itemize} 
\item {\em Homogeneity of space}. If $\tau$ is a translation, then the
region $\tau A$ has the same set of states as $A$.

The function mapping the global state of a system at time 
$t$ to its global state at time $t + T$ commutes with 
all translations.

\item 
{\em Homogeneity of time}. The function mapping the global state 
of a system at time $t$ to its global state at time 
$t + T$ is independent of $t$.

\item 
{\em Bounded density of information}. 
If $A$ is a region of finite size, then the state space of $A$, 
$\Sigma(A)$, is a finite set.

\item 
{\em Bounded velocity of propagation of information}. There exists a
constant $T$ such that for any region $A$, any point in time $t$, the
state of $A$ at time $t + T$, $\rho(A,t+T)$ depends only on 
$\rho(A',t)$, with $A'$ the region of radius $1$ around $A$.  

\item {\em Quiescence}. For each region $A$, there exists a
canonical state $q_A$ called {\em the quiescent sate}. If a
region $A$ is in the quiescent state $q_A$, then the state of any
subset $B$ of $A$ is the quiescent state $q_B$.
At the origin, all the space, except a region of
finite size, is quiescent and the global evolution preserves this
fact. 

Consider a region $A$ that partitions into two regions $B$ and $C$. 
We know that if $A$ is quiescent, then both $B$ and $C$ are quiescent. 
Conversely, as the state of $A$ is determined by the state of $B$ and $C$,
if $B$ and $C$ are quiescent then so is $A$.

Combined with the {\em bounded velocity of information} and the {\em
homogeneity of space}, the {\em quiescence} hypothesis implies that if
the region $A'$ around $A$ is quiescent at time $t$, then so is $A$ at
time $t + T$.  \end{itemize}

\begin{definition}
Let $K$ be the one-to-one mapping from
${\mathbb N}^2$ to ${\mathbb N}$, defined by $K(n,p) =
(n+p)(n+p+1)/2+n$ and $;$ be the one-to-one mapping from
${\mathbb N}^2$ to ${\mathbb N} \setminus \{0\}$, defined by $n;p =
K(n,p)+1$. Let $N$ be the one-to-one mapping from ${\mathbb Z}$ 
to ${\mathbb N}$ defined by $N(x) = 2x$ if $x \geq 0$ and 
$N(x) = - 2x - 1$ if $x < 0$. 
Let $[.]$ be the one-to-one mapping from ${\mathbb Z}^3$ to 
${\mathbb N}$ defined by $[n,n',n'']= K(N(n),K(N'(n),N(n'')))$.

Let $\ulcorner . \urcorner$ be the one-to-one mapping from 
the set of finite sequences of natural numbers to ${\mathbb N}$
defined by 
$\ulcorner j_1, \ldots, j_{l-1}, j_l\urcorner = 
(j_1 ; \ldots (j_{l-1}; (j_l; 0))
\ldots)$. 
\end{definition}

\begin{theorem}[Gandy]\label{prop:gandy} Under the setting and hypotheses
above and given the initial
global state, the function mapping the natural number $k$ to the
global state at time $kT$ is a computable function.  
\end{theorem}

\proof{
$[Partition].$~ 
In the tridimensional physical space, we chose a coordinate 
system $O, {\bf i}, {\bf j}, {\bf k}$
and we consider a partition of
the space into cubic cells of the form $[x,x+1) \times [y,y+1) \times
[z,z+1)$ with $x, y, z \in {\mathbb Z}$.
We also consider a set of translations $\mathcal{T}$ described by vectors of 
the form $x {\bf i} + y {\bf j} + z {\bf k}$
with $x, y, z \in {\mathbb Z}$. 
Each cell and each translation is referenced by a triple
of integers $\langle x, y, z \rangle$ and can be indexed by the number
$[x,y,z]$.

This choice of partition, indexing
and set of translations, is one amongst many that respect the
following properties: 
\begin{itemize}
\item if $C$ is a cell and $\tau$ is a translation in $\mathcal{T}$,
then $\tau C$ is also a cell;
\item conversely, if $C$ and $D$ are two
cells, then there exists a translation $\tau$ of ${\cal T}$, such
that $D = \tau C$;
\item the index of the cell $\tau C$ 
can be computed from the index of $\tau$ and that of $C$;
\item there exists a finite number of translations 
$\sigma_1$,\ldots, $\sigma_r$ such that 
the cells intersecting 
the area of radius 1 around a cell $C$ are 
$\sigma_1 C$,\ldots, $\sigma_r C$.
\end{itemize}

\medskip

$[\Sigma(A)=S].$~ Call $\Sigma(A)$ the set of states of a cell $A$. As 
each cell is of finite size, and using the {\em bounded density of
information} hypothesis, all the $\Sigma(A)$ are finite. Using 
the fact that each 
cell can be obtained by a translation from any other and the 
{\em homogeneity of space} hypothesis, the set of states is the
same for each cell, call it $S=\{e_1,\ldots, e_n\}$, with $e_1=q$ 
the quiescent state.

\medskip

$[a= wqqq\ldots].$~ Using the {\em quiescence} hypothesis, at
the origin of time, and at all times, only a finite number of cells are in a
non-quiescent state.

Thus a global state is 
a function from ${{\mathbb Z}}^3$ to $S$, 
associating a state to each cell, that are equal to $q$ almost 
everywhere.
As both cells and states are indexed, a global state $c$
can be represented as an infinite sequence $j$ 
of elements of 
$\{1, ..., n\}$, such that $p = j_k$ if and only if 
$k = [x,y,z]$ and $e_p = c(x,y,z)$. 
The sequence $j$ is equal to $1$ almost everywhere. Thus we can 
also represent the global state $c$ by the natural number 
$a = \ulcorner j' \urcorner$ where 
$j'$ is the shortest sequence such that $j = j' 1 1 1 \ldots$
This natural number $a$ is the {\em index} of the global state $c$.

If $a$ is a global state, we write $a(C)$ for the state of the cell 
(of index) $C$ in the global state (of index) $a$.

\medskip

$[G(t)(a)=G(a)].$~ Call $G(t)$ the function mapping (the index of) the
global state at time $t$ to (the index of) the global state 
at time $t + T$.

Notice that $G(t)$ is a function mapping global states to global states,
$G(t)(a)$ is a global state and $G(t)(a)(C)$ is a cell state. 

Using 
the {\em homogeneity of time} the function $G$ is independent of 
the time $t$, i.e. there exists a function $G$ such that for all $t$
and $a$, $G(t)(a) = G(a)$.

\medskip

$[G(a)(C) = X(C,a(\sigma_1 C), ...,a(\sigma_r C))].$
Using the {\em bounded velocity of propagation of information}, the state 
of each cell $C$ at a time $t + T$ depends only of the state at time $t$ of the
finite number of cells, $\sigma_1 C, \ldots, \sigma_r C$, that intersect the area $A$ around
this cell of radius $1$. 
Thus, there exists a function $X$ such 
that for all $a$ and $C$, 
$G(a)(C) = X(C, a(\sigma_1 C), \ldots, a(\sigma_r C))$.

\medskip

$[X(C,s_1, ...,s_r) = \chi(s_1, ...,s_r)].$
Let $C$ and $D$ two cells and $s_1, ..., s_r$ be elements of $S$. 
Let $\tau$ be a translation such that $\tau(C) = D$ and
$a$ a state such that $a(\sigma_1 D) = s_1, ..., 
a(\sigma_r D) = s_r$. 
Using the {\em homogeneity of space}
hypothesis $G$ commutes with the function $\Delta$ which sends the
content any cell $C$ into that of cell $\tau(C)$. 
Thus, $G(a) \circ \tau = G(a \circ \tau)$, i.e.
$G(a)(\tau(C)) = G (a \circ \tau)(C)$,
i.e. 
$X(D,a(\sigma_1 D), ..., a(\sigma_r D)) = 
 X(C,a(\sigma_1 D), ..., a(\sigma_r D))$,
i.e. $X(D,s_1, ..., s_r) = X(C,s_1, ..., s_r)$. 
Thus, there exists a function 
$\chi$ such that $X(C,s_1, ...,s_r) = \chi(s_1, ...,s_r)$.

\medskip

[Computability] 
As the function $\chi$ is finite, it is computable.
The function $G$ can be reconstructed from $\chi$ with  $G(a)(C) =
\chi(a(\sigma_1 C), \ldots, a(\sigma_r C)$. Thus it is computable.
Let $a_0$ the the initial global state.
As the function $G$ is computable, the function 
$k \mapsto G^k(a_0)$ mapping the natural
number $k$ to the state of the system at time $kT$ is computable.
}

\smallskip

\begin{remark} 
The proof above uses a fixed orthonormal coordinate system to
define the partition. As nothing is assumed about this coordinate
system, any other could have been chosen.
\end{remark}

\section{Necessity}
\label{subsec:classicalcounter}

Each of the hypotheses is necessary for Proposition
\ref{prop:gandy} to hold. Indeed, we will now, in turn, drop one of
these hypotheses whilst continuing to assume the four others, and show
that the Proposition can then be disproved.  Several of the
counter-examples provided here have already been noticed in the
literature \cite{BeggsTucker2006,BeggsTucker2007-1,BeggsTucker2007-2}
and similar examples may have inspired \cite{Gandy}. But it is
useful to list them in a concise fashion; to this end we reuse the
notations of Section \ref{sec:classic}, choose $U$ some
undecidable subset of $\mathbb{N}$, and define $f_U$ as the non 
computable one-to-one function from ${\mathbb N}$ to ${\mathbb N}$ 
mapping the $n^{th}$ element of $U$ to 
$2n$ and the $n^{th}$ element of ${\mathbb N} \setminus U$ to 
$2n + 1$. 

\begin{itemize} 
\item {\em Without homogeneity of space}, the irregularities in space
could be used to encode $U$.

If the state space associated to each cell is translation-invariant but 
the function $G$ is not, for an initial configuration $a$ of
alphabet $S=\{q,0,1\}$, we would not be able to exclude that the
global dynamics $G$ does a ${\sc Not}$ upon the content $a(\tau^i(C))$
of the $i^{th}$ cell $\tau^i(C)$ if and only if $a(\tau^i(C))$ is in $\{0,1\}$
and $U(i)=1$. (More concisely, $G(a)(\tau^i(C))={\sc
Not}^{U(i)}(a(\tau^i(C)))$.) Such a dynamics could be used to compute
$U$ just by setting $a(\tau^i(C))$ to $0$ and reading off
$G(a)(\tau^i(C))$. 

The same trick can be played if the state space associated to each cell is
not translation-invariant, for instance using 
$\Sigma(\tau^i(C))=\{q,2i,2i+1\}$.  

\item {\em Without homogeneity of time}, the irregularities in
behaviour of the dynamics could be used to encode $U$. We would not be
able to exclude that $G(t+iT)(a)(C)={\sc Not}^{U(i)}(a(C))$.  

\item {\em Without bounded density of information}, the dynamics of
each individual cell would be unconstrained. For instance with
$S=\mathbb{N}$, we would not be able to exclude that
$G(a)(C)=f_U(a(C))$.  

\item {\em Without bounded velocity of propagation of information},
the way the dynamics deals with sets of cells is too loose.  For
instance with $S=\{q,0,1\}$, 
we would not be able to exclude a dynamics that 
maps a segment of cells of the form 
$q x 1^i q$ where $x = 0$ or $x = 1$ to 
$q {\sc Not}^{U(i)}(x) 1^i q$. 

\item {\em Without quiescence}, the initial configuration could be
used to encode $U$. Choosing a trivial $G$, if it is given and
uncomputable input, it will obviously yield an uncomputable
output. This hypothesis is just a way to state that the input
configuration is computable.

\end{itemize}

\section{Hypotheses in the quantum case}
\label{sec:quantumhypotheses}

There are of course several criticisms one can make about Gandy's
hypotheses about the physical world, and these hypotheses have indeed
been criticized.

The hypothesis of finite density of information, in
particular, seems inspired by the idea of `quantization' of the state
space, but is in blatant contradiction with Quantum theory. Indeed in
Quantum theory even a system with two degrees of freedom, i.e. the
qubit, has an infinite state space $\{\alpha\ket{0}+\beta\ket{1}\,|\,
|\alpha|^2+|\beta|^2=1\}$. 

The hypothesis of finite velocity of
propagation of information could also, in some particular EPR-paradox
sense, be said to contradict Quantum theory. Notice however that in
the EPR-paradox no `accessible' information can be communicated faster
than the speed of light \cite{Bell}. Similarly, it can be proved that
not more that one bit of `accessible' information can be stored within
a single qubit \cite{Holevo}. Drawing this distinction between the `description' of the quantum states (infinite, non-local) and the information that can actually be accessed about them, hints
towards the quantum version of these hypotheses.

\subsection{Bounded density of information}
\label{subsec:boundeddensity}

{\em Dimension.} 
As we have seen 
the hypothesis that information has a finite 
density cannot be formulated as the fact that the set of states of
a given cell is finite: in the quantum case this set is always infinite.
Yet, this does not mean that the amount of possible
outcomes, when measuring the system, is itself infinite.  Thus, the
bounded density of information principle can be formulated as the fact
that each projective measurement of a finite system, at any given point in 
time, 
may only yield a finite number of possible outcomes.  
This requirement amounts to the fact that the state space of each
cell is a finite-dimensional vector space.
It constitutes a good quantum
alternative of Gandy's formulation of the finite density of
information hypothesis --- one which does not demand that cells be 
actually measured in any way. 

\smallskip

\noindent {\em Scalars.} The field ${\mathbb C}^2$ includes states such $\lambda
\ket{0} + \mu \ket{1}$, where $\lambda$ is a non-computable real
number and $\mu$ any number such that $|\lambda|^2 + |\mu|^2 = 1$, for
instance, $\lambda$ has a $1$ in the $i^{th}$ decimal if the $i^{th}$
Turing Machine halts and a $0$ otherwise. In order to avoid such
scalars, we shall also assume that the state space of each cell is
defined over a finite extension of the field of rationals. Since we
are in discrete-time discrete-space quantum theory, such a restriction
as little consequences: we have all the scalars that can be generated
by a universal set of quantum gates for instance \cite{BoykinGates},
see also \cite{ArrighiLINEAL} for a more in-depth
discussion. Nevertheless, in the continuous picture, this kind of
assumptions are not without consequences, and these are currently
being investigated \cite{ConnesNumbers,BenioffNumbers}.

\subsection{Bounded velocity of propagation of information}\label{subsec:boundedvelocity}

\noindent {\em Entanglement and state of a subsystem.} In the classical case 
we could assume that 
the state of a compound system was simply given by the 
state of each component. In the quantum setting this no longer holds; 
some correlation information needs to be
added. In other words, the state space of two regions is not the
cartesian product of the state space of each region, but its tensor 
product. Actually if we stick to state vectors, knowing the
state vector (e.g. $(|0\rangle \otimes |0\rangle) + (|1\rangle \otimes
|1\rangle)$) of the compound system, we cannot even assign a state vector to 
the first system. In order to do so, we must switch to the density matrix formalism. 
Each state vector $|\psi\rangle$ is then replaced by the pure density
matrix $|\psi\rangle \langle \psi|$ and if $\rho$ is the 
density matrix of a compound system, then we can assign a density matrix to each 
subsystem --- defined as a partial trace of $\rho$. (The partial trace is 
defined by mapping $A \times B$ to $A$ and extending linearly to $A \otimes B\to A$).
Still, knowing the density matrix of each subsystem is again not 
sufficient to reconstruct the state of the compound system.

\smallskip

\noindent {\em Causality plus unitarity implies localizability.}
The above shows how delicate it is to
formalize the {\em bounded velocity of propagation of information}
hypothesis in the quantum setting. The most natural way to do so has
been formalized in \cite{ArrighiLATA} where it was referred
to as `Causality'. It says that: ``There exists a constant $T$ such
that for any region $A$, any point in time $t$, the density matrix
associated to the region $A$ at time $t + T$, $\rho(A,t+T)$ depends
only on $\rho(A',t)$, with  $A'$ the region of radius $1$
around $A$.'' Actually this definition is a rephrase of
the $C^*$-algebra formulation found in \cite{SchumacherWerner}, which
itself stems from quantum field theoretical approaches to enforcing
causality \cite{Buchholz}.\\ The difficulty of this axiomatic formalization
of the bounded velocity of propagation of information in the quantum
case, is that it is rather non-constructive. As we have explained, 
it is no longer the case that because we know that
$\rho(A,t+T)$ is a local function $f_A$ of $\rho(A',t)$, and
$\rho(B,t+T)$ is a local function $f_B$ of $\rho(B',t)$, then
$\rho(A\cup B,t+T)$ can be reconstructed from $\rho(A'\cup B',t)$ by
means of these two functions.\\ A more constructive approach to
formalizing the bounded velocity of propagation of information in the
quantum case would be to, instead, state that the global evolution is
`localizable' \cite{Beckman,EggelingSchlingemannWerner,SchumacherWestmoreland,ArrighiUCAUSAL}, meaning that the global evolution is
implementable by local mechanisms, each of them physically
acceptable. Here this would say that the global evolution $G$ is in
fact quantum circuit of local gates with infinite width but finite depth. The
disadvantage of this approach in the context of this paper is that
this is a strong supposition to make.\\ Fortunately, in
\cite{ArrighiUCAUSAL,ArrighiJCSS}, the two approaches where shown to be
equivalent. Hence we only need to suppose the former, axiomatic
version of the hypothesis. 

\subsection{Quiescence}
\label{subsec:quiescence}

The quiescence hypothesis remains the same as in the classical case, except that we 
need to assume that the quiescent states are pure states, in order to 
obtain that a region $A$ that partitions into two regions $B$ and
$C$, is quiescent if and only if both $B$ and $C$ are quiescent.

\subsection{Overall}

\begin{itemize}
\item 
{\em Homogeneity of space}. 
As in the classical case.

\item 
{\em Homogeneity of time}.
As in the classical case.

\item {\em Bounded density of information}. The state space of each
finite region is a finite-dimensional vector space over a finite
extension of the rationals.

\item 
{\em Bounded velocity of propagation of information}. There exists a constant $T$ such that for any region $A$, any point in time
$t$, the density matrix associated to $A$ at time $t + T$, 
$\rho(A,t+T)$ depends only on $\rho(A',t)$, with  $A'$ the region of radius $1$
around $A$.

\item 
{\em Quiescence}. 
For each region $A$ of space, there exists a canonical
pure state vector $\ket{q}^A$ called {\em the quiescent sate}. 
If a region $A$ is in the quiescent state $\ket{q}^A$, then the state of any subset
$B$ of $A$ is the quiescent state $\ket{q}^B$. 
At the origin, all the space, except a region of
finite size, is quiescent. The global evolution preserves this
fact. 

\item 
{\em Unitarity}. The global evolution from any point in time $t$ to any other $t+T$ is a unitary operator.

\end{itemize}

\section{A quantum version of Gandy's theorem}

We first define the space that will be used to describe a global state of the 
system. 

\begin{definition}[The Fock space ${\mathcal H}$]
Let $K$ be a finite extension of the field of rationals
and $\Sigma$ be a finite-dimensional $K$-vector space of basis 
$\{{\bf e}_1, \ldots, {\bf e}_n\}$. 
We also write $\ket{q}$ for the vector ${\bf e}_1$.

Let $\mathcal{C}$ be the set of configurations, i.e. functions from 
${\mathbb Z}^3$ to $\{{\bf e}_1, \ldots, {\bf e}_n\}$ that are equal to 
$\ket{q}$, i.e. ${\bf e}_1$, almost everywhere. As both cells and base vectors are indexed, a configuration can be 
represented as an infinite sequence $j$ of elements of $\{1, ..., n\}$, 
such that $j_k = p$ if and only if $k$ is the index of the triple 
$\langle x,y,z \rangle$ and $c(x,y,z)={\bf e}_p$. 
We write this configuration ${\bf e}_j$.
The sequence $j$ is equal to $1$ almost everywhere. 
Thus, we can also 
represent
the configuration ${\bf e}_j$ by the natural number 
$a$, the index of the shortest sequence $j'$ 
such that $j = j' 1 1 1 \ldots$
This natural number is the {\em index} of the configuration ${\bf e}_j$.

The vector space ${\mathcal H}$ is the $K$-vector space of formal
linear combinations of elements of $\mathcal{C}$. The set
$\mathcal{C}$ is an orthonormal basis of this space. 

We define an operation $\otimes$ from $\Sigma \times {\mathcal H}$ to 
${\mathcal H}$ as the bilinear operation mapping the vector 
${\bf e}_i$ and the configuration ${\bf e}_{j_1, j_2, \ldots}$ to 
the configuration ${\bf e}_{i, j_1, j_2, \ldots}$
\end{definition}

As the space ${\cal H}$ is of countable dimension over a countable
field, it is itself countable and can be indexed, for instance, 
we can index the vector 
$\lambda_1 {\bf e}_{j^{1}} + \ldots + \lambda_k {\bf e}_{j^{k}}$
by the number 
$\ulcorner s(\lambda_1), \ulcorner j^{1} \urcorner, \ldots,
s(\lambda_k), \ulcorner j^{k} \urcorner \urcorner$ .
However, unlike
in the classical case where only finite sequences of natural numbers
were indexed (and we know that the choice of an indexing is immaterial in this case, 
provided that list operations remain computable via the chosen indexing)
, we need to be more cautious when indexing
the space ${\cal H}$. We use the fact that, as the structure $\langle
K, \Sigma, {\mathcal H}, +, \times, +, ., +, ., \otimes \rangle$ is
finitely generated relatively to the field $\langle K, +, \times
\rangle$ \cite{ArrighiCIE}, the choice of an indexing for ${\cal H}$
is again immaterial, provided the indexing is 
chosen in such a way that the operations of the structure are
computable.

In the classical case, an important role was played in the proof by the
fact that finite functions are computable. The analogue in the quantum case
is the computability of local linear maps.

\begin{definition}[Local linear map]
A linear map $\phi$ from ${\cal H}$ to ${\cal H}$ is said to be 
{\em local} if there exists an integer $p$ and an linear map $L$ 
from $\Sigma^{\otimes p}$ to $\Sigma^{\otimes p}$, 
such that for any finite sequence $i_1, \ldots, i_p$ of length $p$ and 
infinite sequence $j_1, j_2, \ldots$ equal to $1$ almost 
everywhere,
$$\phi ({\bf e}_{i_1} \otimes \ldots \otimes {\bf e}_{i_p} \otimes 
{\bf e}_{j_1} \otimes {\bf e}_{j_2} \otimes \ldots)
= L({\bf e}_{i_1} \otimes \ldots \otimes {\bf e}_{i_p})
\otimes {\bf e}_{j_1} \otimes {\bf e}_{j_2} \otimes \ldots$$
\end{definition}

\begin{proposition}\label{prop:infinitelinop}
If $\phi$ is a local linear map from $\mathcal{H}$ to $\mathcal{H}$,
then $\phi$ is computable.
\end{proposition}

\proof{Let ${\bf u}$ be an arbitrary vector of ${\cal H}$ and $\lambda_{i,j}$ 
its coordinates 
$${\bf u} = \sum_{i,j} \lambda_{i,j} ({\bf e}_i \otimes {\bf e}_j)$$
Let $J$ be the finite set of infinite sequences $j$ such that 
$\lambda_{i,j}$ is different from zero for some $i$. Then
$$\phi({\bf u}) = 
\sum_{i,j}\lambda_{i,j} ((\sum_{i'} L_{i',i} {\bf e}_{i'}) \otimes {\bf e}_j) 
= \sum_{i',j} (\sum_i L_{i',i} \lambda_{i,j}) ({\bf e}_{i'}\otimes {\bf e}_j)$$
and the coordinate of the vector $\phi({\bf u})$ along the base 
vector ${\bf e}_{i'} \otimes {\bf e}_{j'}$ is 
$\sum_{i} L_{i',i}\lambda_{i,j'}$.
This coordinate is $0$ when $j'$ is not an element of $J$.

If the vector ${\bf u}$ is provided as an index 
$$\ulcorner s(\lambda_1), \ulcorner i^1 j^1 \urcorner, 
\ldots, s(\lambda_k), \ulcorner i^k j^k \urcorner \urcorner$$
then an index of the vector $\phi({\bf u})$ is 
$$\ulcorner s(\sum_{i} L_{{i'}^1,i} \lambda_{i,{j'}^1}), 
\ulcorner {i'}^1 {j'}^1 \urcorner, \ldots,
s(\sum_{i} L_{{i'}^{k'},i} \lambda_{i,{j'}^{k'}}), 
\ulcorner i^{k'} j^{k'} \urcorner \urcorner$$
where ${i'}^1 {j'}^1, \ldots, {i'}^{k'} {j'}^{k'}$ are all the sequences 
where $i'$ is a finite sequence of length $p$ and $j'$ an element of $J$.

The function mapping 
$$\ulcorner s(\lambda_1), \ulcorner i^1 j^1 \urcorner, 
\ldots, s(\lambda_k), \ulcorner i^k j^k \urcorner \urcorner$$
to $$\ulcorner s(\sum_{i} L_{{i'}^1,i} \lambda_{i,{j'}^1}), 
\ulcorner {i'}^1 {j'}^1 \urcorner, \ldots,
s(\sum_{i} L_{{i'}^{k'},i} \lambda_{i,{j'}^{k'}}), 
\ulcorner i^{k'} j^{k'} \urcorner \urcorner$$
is computable, thus the linear map $\phi$ is computable.}

\begin{theorem}
\label{th:mainresult} 
Under the setting and
hypotheses of Section \ref{sec:quantumhypotheses} and given the
initial global state, the function $G$ mapping the natural number $k$
to the global state at time $kT$ is a computable function relatively to
some indexing of the state space. 
\end{theorem}

\proof{$[Partition].$~ We consider the same partition of space into cells as in the classical case.
 
\smallskip

$[\Sigma(A)=\Sigma].$~ Call $\Sigma(A)$ the set of states of the cell $A$. As each cell is of finite size, and using the {\em finite density of information} hypothesis, all the $\Sigma(A)$ are finite-dimensional vector space over 
a field $K$ that is finite extension of the field of rationals --- or more precisely the set of density matrices upon them. 
Using the fact that each cell can be obtained by a translation from
any other and the {\em homogeneity of space} hypothesis, the set of states 
is the same for each cell. Call it $\Sigma$, and choose a basis $\{{\bf e}_1, \ldots, {\bf e}_n\}$, with ${\bf e}_1=\ket{q}$, the quiescent state of the cell.

\smallskip

$[\ket{\psi}= \ket{\phi}\ket{qq\ldots}].$~ Using the {\em
Quiescence} hypothesis, at the origin of time, and at all times, 
only a finite number of
cells are in a non-quiescent state. 
Thus, we can identify the state space with the space ${\mathcal H}$.
We call $\ket{\psi_0}$ the initial global state. 

\smallskip

$[G(t)(\ket{\psi})=G(\ket{\psi})].$~ Call $G(t)$ the function mapping the state of the whole system at 
time $t$ to the state of the whole system at time $t + T$. Using 
the {\em homogeneity of time} this function is independent of 
the time $t$, i.e. there exists a function $G$ such that for all $t$
and $\ket{\psi}$, $G(t)(\ket{\psi}) = G(\ket{\psi})$.

\smallskip

$[G= \prod Swap \prod K_{C}].$~ Using the {\em bounded velocity of propagation of information}, the state 
of each cell $C$ at a time $t + T$ depends only on the state at time $t$ of the
finite number of cells, $\sigma_1 C, \ldots, \sigma_r C$, 
that intersect the area of radius one around
this cell.
This property may be called the {\em causality} of $G$.

As the operator $G$ is both causal and unitary (by the {\em unitarity} 
hypothesis), we can apply the 
Arrighi-Nesme-Werner theorem \cite{ArrighiUCAUSAL,ArrighiJCSS}. 
This theorem requires that each cell $C$ of state space $\Sigma$
be equipped with an ancillary cell $C'$ of state space $\Sigma$. If we
denote $Swap_{C}$ the Swap gate between cell $C$ and cell $C'$, we have
that
$$G=(\prod_{C\in{\mathbb Z}^3} Swap_{C})(\prod_{C\in{\mathbb Z}^3} K_C)$$
with $K_C=G Swap_C G^\dagger$. Notice that the $K_C$ commute with one
another and act only upon $C', \sigma_1 C$, \ldots, $\sigma_r C$, 
see \cite{ArrighiUCAUSAL,ArrighiJCSS} for details. 

\smallskip

$[G= \prod Swap \prod K].$~ Let $\tau$ be any translation. Using the
properties of the translation, the cells that intersect the area
around $\tau C$ of radius is $1$ are the cells $\sigma_1 \tau C$,
\ldots, $\sigma_r \tau C$. 
Using the {\em homogeneity of space}
hypothesis $G$ commutes with the function $\Delta$ which sends the
content any cell $C$ into that of cell $\tau C$. 
Hence
$\Delta^\dagger G^\dagger=G^\dagger \Delta^\dagger$, i.e. $G^\dagger$ also commutes with translations.  Moreover $\Delta
Swap_{C} = Swap_{\tau C} \Delta$.  Hence $\Delta K_C=\Delta G Swap_C
G^\dagger= G \Delta Swap_C G^\dagger = Swap_{\tau C} \Delta G^\dagger
= Swap_{\tau C} G^\dagger \Delta = K_{\tau C} \Delta$. Therefore
$\Delta K_{C} =K_{\tau C} \Delta$. In other words each $K_C$ is a
fixed, local unitary operator $K$ applied upon $C', \sigma_1 C$,
\ldots, $\sigma_r C$. In the same way that each $Swap_C$ is a fixed,
local unitary operator $Swap$ applied upon $C,C'$. Each $K$ and $Swap$
being local, they are therefore computable by Proposition
\ref{prop:infinitelinop}.

\smallskip

$[G= \prod_F Swap \prod_F K].$
Notice also that $K_C|\ldots qqqq\ldots\rangle = |\ldots
qqqq\ldots\rangle$ because of the {\em quiescence} hypothesis, therefore $G$ and hence $G^\dagger$ preserve quiescence, and so does  $Swap_C$.
Hence $K$ applied upon quiescent cells $\sigma_1 C$, \ldots, $\sigma_r C$ 
leaves them quiescent. 
For any state $\ket{\psi}$, only a finite number of cells are in a
non-quiescent state. Let us call $A_{\ket{\psi}}$ this finite region, and $A'_{\ket{\psi}}$ the region around $A$, which is also finite.
Therefore at this time step we have 
$$G\ket{\psi}=G_{A'_{\ket{\psi}}}\ket{\psi}=(\prod_{C\in A'_{\ket{\psi}}}Swap)(\prod_{C\in A'_{\ket{\psi}}} K)\ket{\psi}.$$
This describes an algorithm for computing $G$:\\
- compute $A'_{\ket{\psi}}$ from the index of $\ket{\psi}$;\\
- apply $K$ at each $C$ in $A'_{\ket{\psi}}$;\\
- apply $Swap$ at each $C$ in $A'_{\ket{\psi}}$.\\\ 
Hence the function mapping $k$ to $G^k\ket{\psi_0}$ is computable.}

\section{Necessity.}
\label{sec:qcomments}

Again it is the case that each of the hypotheses
are necessary for Theorem \ref{th:mainresult} to hold. The
counter-examples we have provided in the classical case (See
Section \ref{subsec:classicalcounter}) have been chosen to that
they would also apply in the quantum setting, hence they justify
everything that is left of the classical-case hypotheses within the
quantum-case hypotheses. But there remains some differences:
\begin{itemize} 
\item Within the {\em bounded density of information}
hypothesis in the quantum case, the counter-example we have provided
does show that the state space of each cell needs to be a
finite-dimensional vector spaces. But it does not explain why the
scalars ought to be a finite extension of the rationals. Actually,
there is some degree of freedom as to what kind of scalars should be
allowed, but these should definitely stay within the computable complex
numbers $\tilde{\mathbb{C}}$. Indeed, following the argument given by
\cite{NielsenComputability}, consider the unitary transformation $N$
which maps $\ket{p}$ into $\ket{q}$, $\ket{0}$ into
$u\ket{0}+\sqrt{1-u}\ket{1}$, and $\ket{1}$ into
$\sqrt{1-u}\ket{0}-u\ket{1}$, where $u$ is some uncomputable complex
number of modulus less than one. Let $G=\bigotimes N$, repeated
measurements of the qubits within each cell yield a probabilistic
procedure for approximating $u$, which again is beyond the
computational power of both a deterministic and a probabilistic Turing
machine.

\item On the necessity of the {\em unitarity hypothesis}, it could be argued
that is placed there just in order to be conform with quantum theory
--- and not for the sake of obtaining a computability result. We could
end our discussion here, but on the other hand, it is well-known that
standard quantum theory can be extended to opens systems by allowing more general randomised unitary
evolutions, namely quantum operations (also referred to as
superoperators or TPCP-maps). If we were to allow this extension
however, Proposition \ref{th:mainresult} would no longer hold. In
order to see this, all one needs to know about quantum operations is
that they include probabilistic, classical evolutions. So, let us go
back to the classical setting and suppose that $G$ can now be a
stochastic map. Again take $U$ and undecidable subset of $\mathbb{N}$;
this time the correlations produced by $G$ can be used to encode
$U$. That is we would not be able to exclude that $S=\{q,0,1\}$, and
$G(a)(C)$ equals $q$ if $a(C)=q$, $1$ if $a(C)=1$, and the probability
distribution $\{(1/2,0),(1/2,1)\}$ is $a(C)=0$, with the added
condition that those probability distributions are correlated with one
another if and only if those two initial zeroes where separated by a
distance $i$ cells, and $U(i)=1$. Such dynamics would yield a
probabilistic procedure for computing $U$, just by setting $a$ to
$\ldots qq0(1)^i0qq\ldots$ and then measuring whether the images of
the zeroes are correlated or not. \end{itemize} (This counter-example
does satisfy the {\em bounded velocity of information} hypothesis, but
it does lead to the impression that the {\em bounded velocity of
information} hypothesis that has become too weak in the presence of
quantum operations, due to the lack of a `Unitarity plus causality
implies localizability' theorem as in \cite{ArrighiUCAUSAL,ArrighiJCSS} valid for
quantum operations. But reinforcing notions of causality to account
for quantum operations \cite{Beckman,EggelingSchlingemannWerner,SchumacherWestmoreland} or even just probabilistic
evolutions is renowned to be a difficult topic
\cite{Henson}.)

\section{Conclusion}

{\em Summary.} We have given a quantum version of Gandy's theorem. Namely, assuming only homogeneity of (euclidean) space and time, bounded density of space, bounded velocity of propagation of information, quiescence and unitarity, we have shown that the evolution of a quantum system is computable. Besides the classical version of the theorem \cite{Gandy,CopelandShagrir,SiegByrnes}, there were two key ingredients to this extension.\\ First of all, Quantum theory is about vector spaces, and its evolutions are functions over these vector spaces. Therefore, we needed a `stable' notion of what is means to be computable in this context, a theory provided in \cite{ArrighiCIE}. Our Proposition \ref{prop:infinitelinop} states that local linear operators are computable in this sense; this constitutes an interesting addendum to the theory.\\
Secondly, Quantum theory is about tensor products of vector spaces, i.e. quantum systems are not just put aside but may be entangled. Therefore, whereas causality (i.e.  bounded velocity of propagation of information) immediately provides a local transition function in the classical setting, of which the global evolution is composition, the counterpart is harder to obtain in the quantum setting. For this we have had to resort to the `Unitarity plus causality implies localizability' result provided in \cite{ArrighiUCAUSAL,ArrighiJCSS}. In a sense our Theorem could also be seen as taking this result further, by stating that `Unitarity plus causality implies computability'.

{\em Future work.} This result clarifies when it is the case that
Quantum theory evolutions could break the physical Church-Turing
thesis or not; a series of examples shows that it suffices that one of
the above hypotheses be dropped. This draws the line between the
positive result of \cite{Deutsch} and the negative results of
\cite{NielsenComputability,Kieu,NielsenMore}. Because these
hypotheses are physically motivated, this is a step along Nielsen's
programme of a computable Quantum theory. Further work could be done
along this direction by introducing a notion of `reasonable
measurement' \cite{Peres}, or investigating the continuous-time picture as started by \cite{SmithComputability,WernerComputability}.
Prior to that however this work raises
deeper questions: Is the bounded density of information really
compatible with modern physics? For instance, can we really divide up
space into pieces under this condition, without then breaking further
symmetries such as isotropy? 

{\em Bigger picture.} The question of the robustness of the physical
Church-Turing thesis is certainly intriguing; but it is hard to
refute, and fundamentally contingent upon the underlying physical
theory that one is willing to consider. For instance in the General
Relativity context a series of paper explain how `hypercomputation'
might be possible in the presence of certain space-times
\cite{Hogarth,EtesiNemeti}. Beyond this sole question however, we find
that it essential to exhibit the formal relationships that exist
between the important hypotheses that we can make about our
world. Even if some of these hypotheses cannot be refuted, whenever
some can be related to one another at the theoretical level, then one
can exclude the inconsistent scenarios.

\section*{Acknowledgments}
We are particularly thankful to Vincent Nesme and Reinhard Werner, since their paper with one of the authors \cite{ArrighiUCAUSAL,ArrighiJCSS} contains a rather central ingredient to our proof. We would like to thank Alexei Grimbaum and Zizhu Wang for several comments, as well as Simon Perdrix, Mehdi Mhalla and Philippe Jorrand.

\bibliography{biblio}

\bibliographystyle{plain}

\end{document}